\documentclass[%
 reprint,
superscriptaddress,
 amsmath,amssymb,
 aps,
floatfix,
]{revtex4-2}

\usepackage{graphicx}
\usepackage{dcolumn}
\usepackage{bm}
\usepackage{tabularx}

\begin{document}

\preprint{Investigation of Planckian behavior in a high-conductivity oxide: PdCrO$_2$}

\title{Investigation of Planckian behavior in a high-conductivity oxide: PdCrO$_2$}

\author{Elina Zhakina}
 \email{elina.zhakina@cpfs.mpg.de}
\affiliation{Max Planck Institute for Chemical Physics of Solids, Nöthnitzer Straße 40, 01187 Dresden, Germany}

\author{Ramzy Daou}
\affiliation{Laboratoire de Cristallographie et Sciences des Matériaux (CRISMAT), Normandie Université, UMR6508 CNRS, ENSICAEN, UNICAEN, 14000 Caen, France}

\author{Antoine Maignan}
\affiliation{Laboratoire de Cristallographie et Sciences des Matériaux (CRISMAT), Normandie Université, UMR6508 CNRS, ENSICAEN, UNICAEN, 14000 Caen, France}

\author{Philippa H. McGuinness}
\affiliation{Max Planck Institute for Chemical Physics of Solids, Nöthnitzer Straße 40, 01187 Dresden, Germany}

\author{Markus König}
\affiliation{Max Planck Institute for Chemical Physics of Solids, Nöthnitzer Straße 40, 01187 Dresden, Germany}

\author{Helge Rosner}
\affiliation{Max Planck Institute for Chemical Physics of Solids, Nöthnitzer Straße 40, 01187 Dresden, Germany}

\author{Seo-Jin Kim}
\affiliation{Max Planck Institute for Chemical Physics of Solids, Nöthnitzer Straße 40, 01187 Dresden, Germany}

\author{Seunghyun Khim}
\affiliation{Max Planck Institute for Chemical Physics of Solids, Nöthnitzer Straße 40, 01187 Dresden, Germany}

\author{Romain Grasset}
\affiliation{Laboratoire des Solides Irradis, CEA/DRF/IRAMIS, Ecole Polytechnique,
CNRS, Institut Polytechnique de Paris, F-91128 Palaiseau, France}

\author{Marcin Konczykowski}
\affiliation{Laboratoire des Solides Irradis, CEA/DRF/IRAMIS, Ecole Polytechnique,
CNRS, Institut Polytechnique de Paris, F-91128 Palaiseau, France}

\author{Evyatar Tulipman}
\affiliation{Department of Condensed Matter Physics, Weizmann Institute of Science, Rehovot, 76100, Israel}

\author{Juan Felipe Mendez-Valderrama}
\affiliation{Department of Physics, Cornell University, Ithaca, New York 14853, USA.}

\author{Debanjan Chowdhury}
\affiliation{Department of Physics, Cornell University, Ithaca, New York 14853, USA.}

\author{Erez Berg}
\affiliation{Department of Condensed Matter Physics, Weizmann Institute of Science, Rehovot, 76100, Israel}

\author{Andrew P. Mackenzie}
 \email{andy.mackenzie@cpfs.mpg.de}
\affiliation{Max Planck Institute for Chemical Physics of Solids, Nöthnitzer Straße 40, 01187 Dresden, Germany}
\affiliation{Scottish Universities PhysicsAlliance, School of Physics $\&$ Astronomy, University of St. Andrews, St. Andrews KY16 9SS, United Kingdom}

\date{\today}
\begin{abstract}
The layered delafossite metal PdCrO$_2$ is a natural heterostructure of highly
conductive Pd layers Kondo coupled to localized spins in the adjacent Mott insulating CrO$_2$
layers. At high temperatures $T$ it has a $T$-linear resistivity which is not seen in the isostructural
but non-magnetic PdCoO$_2$. The strength of the Kondo coupling is known, as-grown crystals
are extremely high purity and the Fermi surface is both very simple and experimentally
known. It is therefore an ideal material platform in which to investigate ‘Planckian metal’
physics. We do this by means of controlled introduction of point disorder, measurement of
the thermal conductivity and Lorenz ratio and studying the sources of its high temperature
entropy. The $T$-linear resistivity is seen to be due mainly to elastic scattering and to arise from
a sum of several scattering mechanisms. Remarkably, this sum leads to a scattering rate
within 10$\%$ of the Planckian value of $k_BT/$$\hbar$. 

\end{abstract}

\maketitle
\setlength{\parskip}{0.0cm}
\section{Introduction}
Understanding the physics of so-called `Planckian metals' is a central challenge of modern condensed matter physics.  One of the most notable properties of the high temperature cuprate superconductors, is a d.c. resistivity $\rho$ that depends linearly on temperature ($T$) from just above the superconducting transition temperature ($T_c$) to hundreds of kelvin, in one notable case crossing two decades of temperature \cite{1}. In any single cuprate material, $\rho$ $\sim$ $\rho_0$ + $AT$ (with $\rho_0$ the residual resistivity) when it is tuned to be close to its maximum $T_c$, suggesting a link between the $T$-linear resistivity and the high temperature superconductivity  \cite{2}.  Not only the power law, but also its prefactor $A$ are significant.  Optical conductivity data on cuprates showing $T$-linear resistivity provided evidence of a resistive scattering rate $\frac{1}{\tau_{tr}}$ varying as $k_B$$T$/$\hbar$, a rate postulated to be the highest allowed in a strongly interacting quantum system and termed 'Planckian dissipation' \cite{3,4,5}.  Interest in $T$-linear resistivity in strongly correlated electron systems grew with the discovery of its existence in other systems tuned to the proximity of quantum critical points, including heavy fermion materials, iron-based and organic superconductors and transition metal oxides \cite{6,7,8,9,10}.  Optical conductivity data is sparser in these systems, but analysis of d.c. resistivity data in certain systems for which there was experimental information on the Fermi surfaces and effective masses m* concluded that the observed values of $A$ are such that $\frac{1}{\tau_{tr}}$ = $\alpha$$\frac{k_BT}{\hbar}$  with 0.7 $<$ $\alpha$ $<$ 2.7  \cite{10}.  Similar analyses applied more recently to d.c. transport data from cuprates \cite{11}, twisted bilayer graphene \cite{12, Nat} and twisted transition metal dichalcogenides \cite{Nat2} and doped two-dimensional semiconductors \cite{Andy1} reached similar conclusions, while $T$-linear resistivity was shown to persist to temperatures as low as 10 mK in YbRh$_2$Si$_2$ \cite{13}. 

The analyses described above rely on accurate knowledge of the parameters of the materials in question, and the use of different Fermi surface averages of the Fermi velocity $v_F$  for $\rho$ $\propto$ $\overline{\frac{1}{v_F(k)}}$ and $m^*$ $\overline{v_F(k)}$ \cite{14}.  This issue can be mitigated by having information on all the Fermi surface sheets, but the different ways of averaging within individual sheets can introduce error if $v_F$ is strongly $k$-dependent. It is therefore highly desirable to study the Planckian problem in simple, preferably single-band systems, in which the Fermi velocity is nearly isotropic around the Fermi surface.

A second important issue when considering Planckian dissipation is the origin of the scattering that produces it.  The Planckian rate $\frac{1}{\tau_{tr}}$ $\cong$ $\frac{k_BT}{\hbar}$ has been noted at high temperatures in electron-phonon systems since the time of Peierls \cite{15}, \cite{16} and understood in terms of a temperature-dependent scattering cross-section for quasi-elastic processes.  The fact that it is approximately bounded in real materials has been postulated to be because of a bound on the stability of the metallic state to polaron formation, i.e. a limit on how high electron-phonon coupling can be while still leading to a conventional metal \cite{17}.  In contrast, if the $T$-linear resistivity originates from inelastic scattering of electrons (as in a `marginal Fermi liquid' \cite{18}), then the transport time is more naturally related to the equilibration time \cite{14}.

The issue of the temperature range over which electron-electron or electron-phonon scattering are microscopically responsible for the observed $\rho$ $\sim$ $\rho_0$ + $AT$ is a matter of active debate in the literature \cite{19}, inviting careful investigatin of different material classes in which $T$-linear resistivity is observed.  There is also surprisingly little systematic investigation of the level to which, when there is scattering at the Planckian rate, $A$ and $\rho_0$ are independent.

The material on which this paper is based, PdCrO$_2$, is ideal for careful experimental investigation of the issues outlined above.  It consists of alternating layers of highly conducting Pd, coupled to Mott insulating layers of CrO$_2$ in which Cr$^{3+}$ is thought to be in the high-spin 3d$^3$ configuration with an effective moment corresponding to spin-3/2 \cite{20,21,22,add1,add2,add3,add4}. Its Weiss temperature ($\theta_W$) is approximately 500 K but magnetic order sets in only at $T_N$ = 38 K.   The ratio between its out-of-plane resistivity $\rho_c$ and in-plane resistivity $\rho_{ab}$ is larger than 150 \cite{23}, justifying a 2D approximation when analyzing its properties. In a striking contrast to its non-magnetic sister PdCoO$_2$, $\rho_{ab}$ is $T$-linear from approximately 150 K to at least 500 K \cite{24}.  

PdCrO$_2$ is particularly well-suited to the analysis of this $T$-linear in-plane resistivity for a number of reasons. Consistent with the large value of $\rho_c$/$\rho_{ab}$, the high temperature Fermi surface is a simple cylinder of nearly hexagonal cross-section, with a nearly constant $v_F$$(k)$ in a 2D Brillouin zone and a large carrier density \cite{25,26,27}. This high temperature Fermi surface is almost identical to that of its partner material PdCoO$_2$ in which Co$^{3+}$ is in the non-magnetic low-spin 3d$^6$ configuration, facilitating a quantitative empirical comparison between the properties of the two materials.  For both materials, the residual resistivity is tiny in as-grown crystals, providing a well-controlled starting point for data interpretation.  Finally, the physics of PdCrO$_2$ has been shown to be well described by a low energy Kondo lattice Hamiltonian describing the inter-layer coupling of conduction electrons with the local moments in the Mott insulator layer \cite{27}, providing firm foundations for its theoretical analysis. This combination of properties makes PdCrO$_2$ attractive as one of the materials best suited for a quantitative empirical study of $T$-linear metallic resistivity.

Working with single crystals and cutting micro-devices from them to perform electrical resistivity measurements, we show that $\frac{1}{\tau_{tr}}$ = 0.9$\frac{k_BT}{\hbar}$ in as-grown crystals of PdCrO$_2$, which remains unchanged as $\rho_0$ is increased by a factor of twenty by the deliberate introduction of defects by high-energy electron irradiation. We present thermal conductivity data from PdCrO$_2$, and a comparative heat capacity study of PdCrO$_2$ and PdCoO$_2$, accompanied by density functional theory calculations of the phonon spectra of the two materials.  Finally, we  outline a theoretical proposal, described in detail in a separate paper \cite{theory}, for how the high temperature resistivity could be explained as a consequence of the electron magneto-elastic coupling between the conducting and magnetic layers.

\section{Experiment and density functional calculations}
\subsection{Sample growth}
Single crystals of PdCoO$_2$ and PdCrO$_2$ were grown in quartz tubes via methods discussed in ref. \cite{28} and ref. \cite{29}. For the microstructuring described below we selected platelets $\sim$ 5–20 $\mu$m thick, with lateral dimensions $\sim$ 300–700 $\mu$m.

\subsection{Electron irradiation}
Crucially for the analysis, the scattering centres introduced to the crystal should be point-like defects and no large voids or columnar defects should be created. High-energy electron irradiation is the ideal technique to achieve this type of disorder. The collision kinetics of 2.5 MeV electrons with much heavier atoms allow transmission of enough energy to an atom to displace it from its lattice site, but not enough for the displaced atom to create a significant number of additional defects.  The collisions therefore create an individual vacancy plus an interstitial atom, known as a Frenkel pair. The 2.5 MeV electrons have a large penetration range estimated to be 1.8 mm in delafossite metals, so the probability of any electron undergoing more than one collision in samples a few tens of microns thick is negligible. Further details can be found in ref. \cite{30}. 

The 2.5 MeV electron irradiation was performed at the SIRIUS Pelletron linear accelerator operated by the Laboratoire des Solides Irradiés (LSI) at the École Polytechnique in Palaiseau, France.  During the irradiation, the sample is immersed in a bath of liquid hydrogen at a temperature of $\sim$ 22 K, and we are able to perform an in-situ four-point resistance measurement by pausing at regular intervals and therefore monitor the increase of resistivity as a function of electron dose.   

\subsection{Sample preparation}

\begin{figure}
    \centering
    \includegraphics[width=0.48\textwidth]{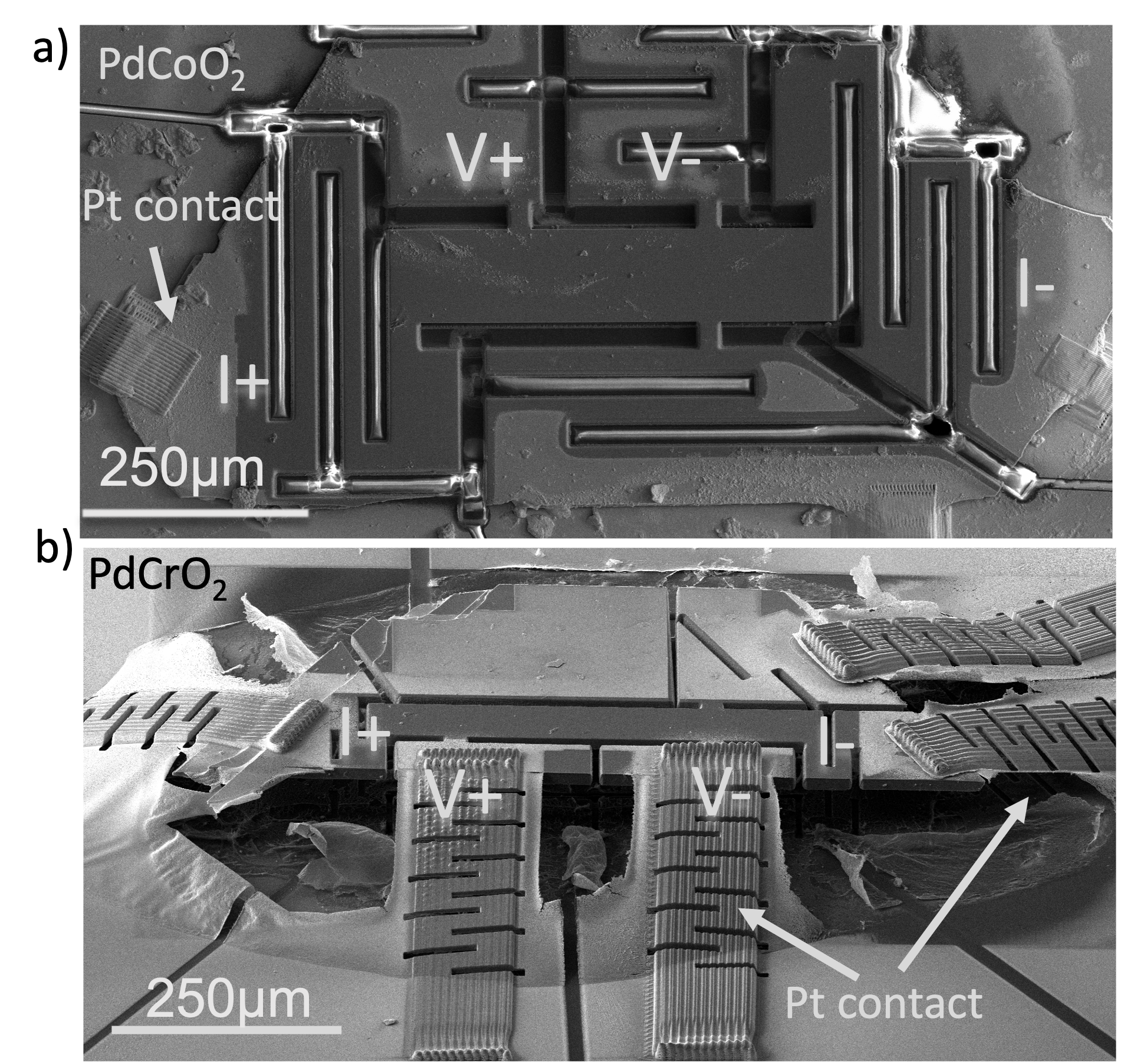}
    \caption{(a) SEM image of a microstructured PdCoO$_2$ sample mounted by ‘epoxy-free’ method. (b) SEM image of the PdCrO$_2$ device mounted using ‘free-standing’ method. The measured regions of two devices appear to be darker in the SEM image because the gold is removed from them. The rest of the devices, Pt contacts, and substrate are covered with sputtered gold.  The Pt contacts are slotted to decrease their spring constant and give the overall device enough mechanical flexibility to avoid the PdCrO$_2$ fracturing during cool-downs.}
    \label{fig1}
\end{figure}

In order to enable a reliable low-temperature resistivity measurement of delafossites, we used focused ion beam sculpting to increase the signal-to-noise ratio and decrease the geometrical uncertainties. The standard method of sample mounting for microstructuring requires a layer of glue to attach the crystal to a substrate, as was done in previous transport studies \cite{31}. However, glue degrades quickly in the electron beam, motivating us to use the 'epoxy-free' method of mounting PdCoO$_2$ described in \cite{30}. A scanning electron micrograph of a PdCoO$_2$ device structured using this method is shown in Fig. \ref{fig1}a. The crystal is held on a substrate by electrostatic forces and pinned by deposited Pt contacts providing a good mechanical connection between the crystal and gold-coated substrate. The second sputtered 150-nm gold layer ensures contact resistance on the order of 1$\Omega$. Unfortunately, the 'epoxy-free' method cannot be used for PdCrO$_2$ as a strong mechanical coupling between the substrate and the crystal causes the device to break during thermal cycling. To decouple the crystal from the substrate, we mounted it in two-component glue and sputtered a 150-nm gold layer. The deposited Pt leads play the role of a bridge connecting PdCrO$_2$ with the substrate. The glue was removed underneath the sample by oxygen plasma etching, leaving crystal 'hanging' on Pt leads, as shown in Fig. \ref{fig1}b. We used a focused ion beam (FIB) to sculpt the PdCrO$_2$ for resistivity measurements as the final step.   

\subsection{Thermal conductivity}

Thermal conductivity was measured using a standard one-heater, two-thermometer technique on the same thin single crystal of PdCrO$_2$ measuring 0.5 x 0.5 x 0.002 mm$^3$ with $\rho_0$ $\sim$ 0.05 $\mu\Omega$cm from ref. \cite{daou}. Temperatures were measured using fine wire thermocouples attached to the sample. The thermal current was measured using a calibrated heat pipe in series with the sample to reduce the error associated with thermal radiation. This results in an uncertainty of $\pm$ 15$\%$ in the absolute values at 300 K, reducing to zero at around 100 K. The systematic uncertainty due to the geometric factor is around $\pm$ 10$\%$. The resistivity of this sample was measured simultaneously using the same contacts, cancelling the geometric uncertainty in the Lorenz ratio.

\subsection{Density functional theory calculations}

Density functional theory (DFT) calculations were performed using the VASP code \cite{VASP1,VASP2,VASP3,VASP4} to optimize the crystal structure and obtain the accurate descriptions of the ground states. The exchange correlation functionals were considered at the local density approximation (LDA) \cite{LDA-CA} and generalized gradient approximation (GGA) \cite{GGA-PBE} levels. A plane-wave cutoff of 600 eV and energy convergence criteria of 10$^{-8}$ eV were used. All geometries were fully relaxed by minimizing the forces on each atom and reducing pressure to zero. To obtain accurate ground state energies, we included the spin-orbit coupling effect, and a 16x16x2 $k$-point mesh was used. The bulk moduli were calculated by fitting the total energies of different volumes using the Birch-Murnaghan method. \cite{BM} 

For the phonon calculations, the frozen phonon method implemented in the Phonopy code \cite{Phonony} was used with 4x4x1 supercell of the fully-relaxed conventional unit cell. We have considered four displacement modes that are allowed by the space group No. 166 (R-3m). A 401x401x401 $q$-point mesh was used to obtain the vibrational heat capacity and entropy.

\section{Results}

\begin{figure}
    \includegraphics[width=0.48\textwidth]{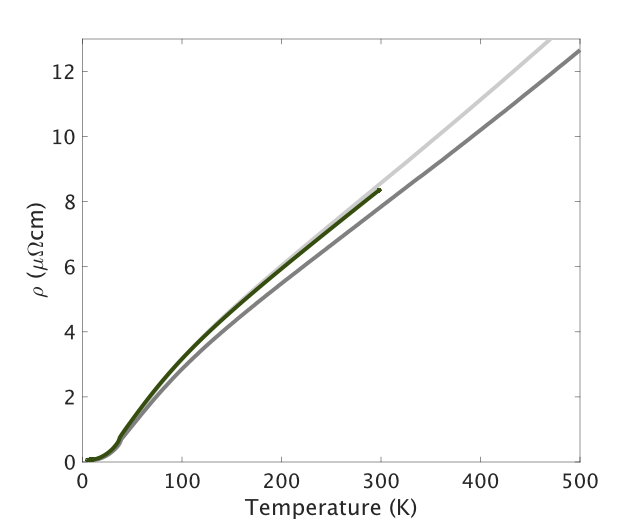}
    \caption{In-plane resistivity data for the PdCrO$_2$ microstructure shown in Fig. \ref{fig1} (green) compared with data from previous work \cite{24} on two single crystals (grey).  The difference between the results from the two crystals is due to inevitable uncertainty in the geometrical factors used to convert resistance to resistivity when working with traditional hand-mounted silver epoxy contacts.  Geometrical errors can be made much smaller with microstructures.  Overall, the agreement between the three measurements is excellent. A $T$-linear resistivity is observed between approximately 150 K and the highest measurement temperature of 500 K.}
    \label{fig2}
\end{figure}

In Fig. \ref{fig2} we show the resistivity measured for the PdCrO$_2$ microstructure shown in Fig. \ref{fig1}, for which the geometrical factors are well-defined, highlighted against previous work going to higher temperatures for which data were taken on bulk single crystals \cite{24}.  The agreement is seen to be excellent, giving confidence that the $T$-linear resistivity seen between 150 K and 300 K in the PdCrO$_2$ microstructure is representative of data going to much higher temperatures of over 500 K.  The gradient of the $T$-linear term is 0.026 $\mu$$\Omega$cm/K. As in all previous reports, there is a sharp drop in resistivity below $T_N$.

De Haas-van Alphen measurements performed on PdCrO$_2$ show that its high temperature Fermi surface (sampled at low temperatures thanks to magnetic breakdown across small gaps in the low temperature antiferromagnetically ordered state) has an average cyclotron mass $m^*$ = 1.55 $m_e$, where $m_e$ is the electron mass \cite{24,32}.  This corresponds to a Fermi-surface averaged Fermi velocity of approximately 7.5 x 10$^5$ ms$^{-1}$.  Angle-resolved photoemission spectra from delafossites are extremely sharp, and give no evidence that this Fermi velocity is temperature-dependent \cite{K1}.  With a Fermi temperature of approximately 25000 K, the $T$-independence of $v_F$ across the range of temperatures relevant to this paper is unsurprising.  In common with observation and calculation on other delafossite Fermi surfaces \cite{31,33,34,35}, angle-resolved photoemission spectroscopy \cite{25,26,27} shows that the variation of the Fermi velocity about its mean is less than 10$\%$ in the two-dimensional Brillouin zone appropriate for PdCrO$_2$. Adopting this two-dimensional approximation appropriate gives 

\begin{multline}
        \sigma(T) = \frac{1}{\rho(T)} = \frac{e^2}{hd}k_Fv_F\tau_{tr}(T)\\ \Rightarrow \frac{1}{\tau_{tr}(T)} = \frac{e^2}{hd}k_Fv_F\rho(T),
\end{multline}

in which e is the electronic charge, $h$ is Planck’s constant, $d$ = 6.03 \AA $ $ is the interlayer spacing and $k_F$ = 0.93 x 10$^{10}$ m$^{-1}$ is the average in-plane Fermi wavevector.  Combining with the measured temperature-dependent $\rho(T)$ = 0.026 $\mu$$\Omega$cm/K and $v_F$ =7.5 x 10$^5$ ms$^{-1}$ gives $\frac{1}{\tau_{tr}T}$ = 1.2 x 10$^{11}$ s$^{-1}$K$^{-1}$ = 0.9$\frac{k_B}{\hbar}$. The favourable Fermi surface parameters summarized above mean that the degree of averaging required for the analysis is small, as is the mass renormalization. This is therefore one of the most reliable estimates of a scattering rate using a quasiparticle analysis that can be performed, so the closeness of the result to the Planckian value is particularly striking.

Our choice of microstructures for the current experiment was motivated by the desire to perform a careful study of the effect of changing the elastic scattering rate of PdCrO$_2$ by adding disorder. Although this can in principle be done by either irradiating or chemically doping different samples \cite{30}, \cite{36,37,38}, the inevitable uncertainties in geometrical factors between those samples is difficult to separate with confidence from the intrinsic effects of disorder.  Our goal was to obtain a precise measurement of the effect of the extra elastic scattering on the $T$-linear term in $\rho_{ab}$, so it was vital to eliminate geometrical uncertainty from the experiment. As described above, the high energy electron irradiation that provides the required point defects places the sample in a harsh environment, so even working with one single crystal mounted with traditional silver paint contacts is risky. If one of those contacts moved or had to be repaired, the change in geometrical factor could ruin the experiment. In a microstructure of the kind shown in Fig. \ref{fig1}, the geometrical factors are determined by the sculpting of the sample, so as long as it does not break during the experiment, even the contacts between the sample and the measurement wires can be repaired, if necessary, without any change to the geometry of the measurement.  

\begin{figure*}
    \centering
    \includegraphics[width=\textwidth,trim=4 4 4 4,clip]{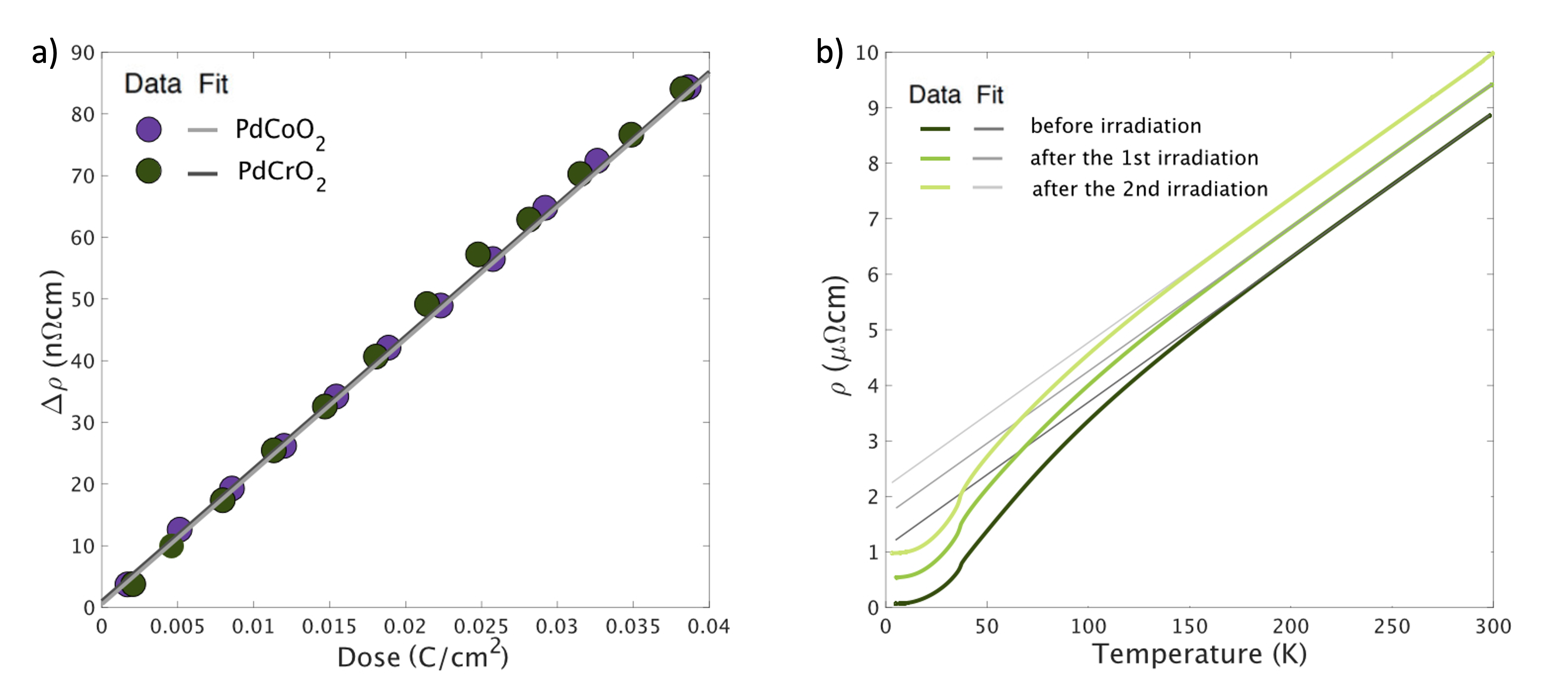}
    \caption{(a) Increase of resistivity as a function of electron dose for two delafossite metals: PdCoO$_2$, and PdCrO$_2$. (b) Change to the in-plane resistivity of the PdCrO$_2$ microstructure shown in Fig. \ref{fig1}, in its as-prepared state and then after two successive exposures to high energy electron radiation with defect concentrations of 0.16$\%$ and 0.29$\%$. The gradient of the $T$-linear resistivity is unchanged, within small experimental errors, although the residual resistivity is increased by over a factor of twenty.}
    \label{fig3}
\end{figure*}

The increase of the resistivity of PdCoO$_2$ and PdCrO$_2$ as a function of electron dose is shown in Fig. \ref{fig3}a. In these two compounds, the dependence of resistivity on dose is linear and has the same slope in the investigated range. The extra resistivity is dominated by defects in the conductive Pd planes, as expected in such two-dimensional systems \cite{30}.  

The results of two long-time irradiations of the same PdCrO$_2$  microstructure are shown in Fig. \ref{fig3}b. In each case the sample was left overnight in the electron beam, and its resistivity measured after transporting it from Paris to Dresden. The second irradiation took place in a different beam time to the first one, so overall the experiment took six months, but it produced a conclusive result: within experimental error the gradient of the $T$-linear resistivity is unchanged in spite of increasing the residual resistivity by over a factor of twenty, from 40 to 900 n$\Omega$cm.

To summarize the conclusions that can be drawn from analysis of Figs. \ref{fig2} and \ref{fig3}: the scattering rate in PdCrO$_2$ can be extracted from the d.c. resistivity with an unusually high degree of confidence because of its simple Fermi surface and favourable quasiparticle parameters. It is 90$\%$ of the Planckian value, and remains unchanged within experimental error even when significant levels of disorder are introduced, substantially changing the levels of elastic disorder scattering.  These are two of the key experimental observations that we report. For the remainder of the paper we consider the scattering mechanisms that could lead to the observed behaviour.

\begin{figure*}
    \centering
    \includegraphics[width=\textwidth,trim=2 2 2 2,clip]{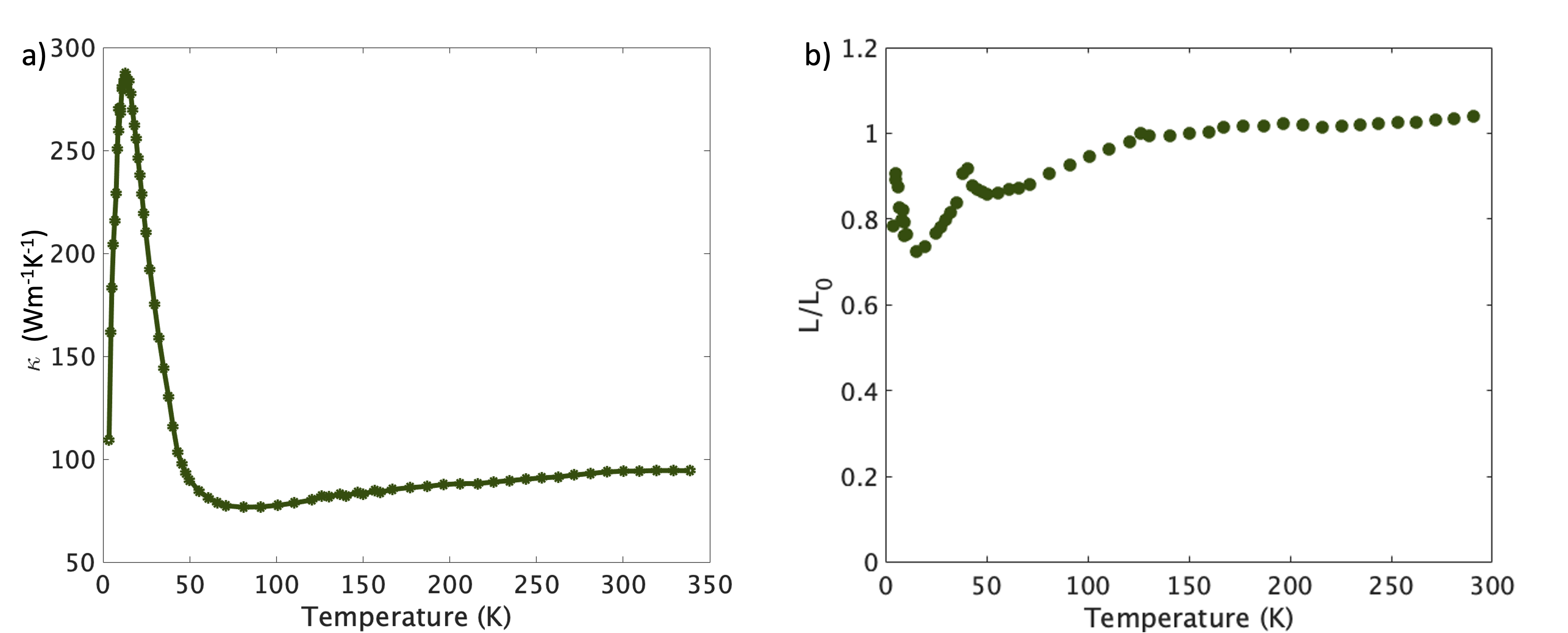}
    \caption{a) Thermal conductivity and b) Lorenz ratio of PdCrO$_2$ as a function of temperature}
    \label{fig4}
\end{figure*}

First, we examine the thermal conductivity $\kappa$, for which data are presented in Fig. \ref{fig4}a.  As expected of a high-conductivity metal, it is large, reaching nearly 300 W/Km at its peak value. It rises with increasing temperature rapidly below $T_N$, qualitatively consistent with the behaviour of the electrical conductivity. Quantitative comparison of electrical and thermal conductivity comes from calculating the Lorenz ratio $L$ = $\frac{\kappa\rho}{T}$. In systems for which electrons dominate heat conduction, and the same average scattering rate determines both the thermal and electrical conductivity, $L$ = $L_0$ = $\frac{\pi^2}{3}$$(\frac{k_B}{e})^2$. The condition for this is that the scattering be quasi-elastic, i.e. that the average energy relaxation rate due to the scattering is much smaller than the average momentum relaxation rate. This is satisfied in any metal at sufficiently low temperatures in the regime in which impurity scattering dominates, and in conventional metals at high temperatures in the regime in which electron-phonon scattering dominates.       

\begin{figure*}
    \centering
    \includegraphics[width=0.95\textwidth]{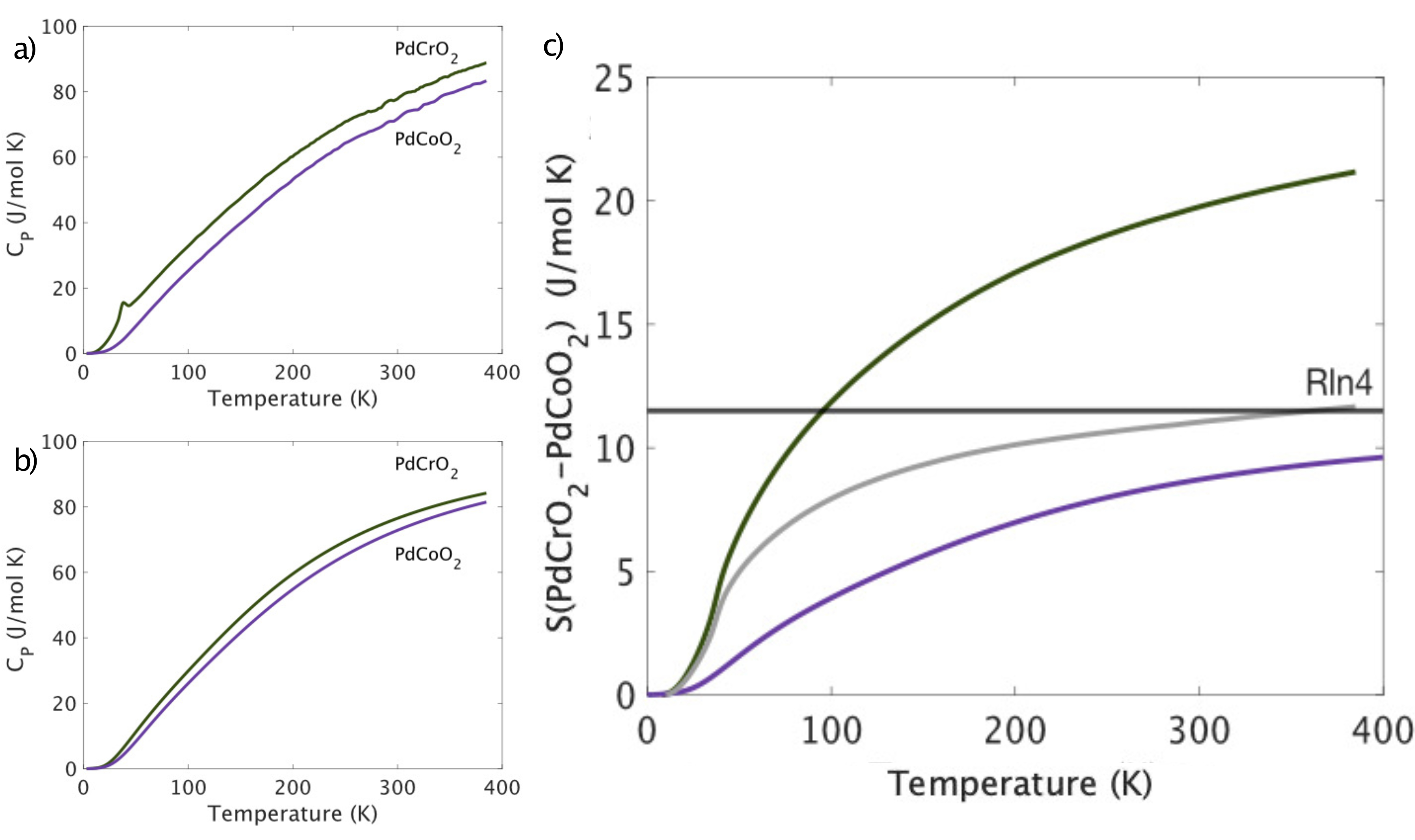}
    \caption{a) The total heat capacity of magnetic PdCrO$_2$ and non-magnetic PdCoO$_2$ as a function of temperature up to 390 K.  b) Density functional calculation for the vibrational heat capacity for non-magnetic PdCoO$_2$ and spin-polarized PdCrO$_2$. c) The entropy difference between the two compounds over the same range of temperatures (green), split into two components: the phonon part (purple) deduced from the calculated data of panel b) (see also Fig. \ref{s3}) and the remainder, which we attribute to magnetic entropy not included in the density functional calculations.  The latter comes very close to the expected value of Rln4 by 390K.}
    \label{fig5}
\end{figure*}

In PdCrO$_2$, $L$ is seen to be within 5$\%$ of $L_0$ for all measured temperatures above 100 K. Similar behaviour is seen in conventional metals, and attributed to electron-phonon scattering. In PdCrO$_2$, however, a body of evidence points to a more complex situation. Firstly, the sharp drop of resistivity below $T_N$ hints at rather strong magnetic scattering which is frozen out as the magnetically ordered state is entered. The existence of the non-magnetic sister compound PdCoO$_2$ gives the opportunity for a complementary estimate of the strength of electron-phonon scattering in PdCrO$_2$. The two compounds have very similar total atomic mass, so their phonon spectra would be expected to be similar; indeed, direct observation shows similar optic phonon frequencies in the two compounds \cite{39,40}.  However, measurement of the heat capacity reveals a fairly large difference between the two, extending to our highest temperature of measurement [Fig. \ref{fig5}a]. Although a magnetic contribution to the heat capacity is expected, and observed \cite{30} in PdCrO$_2$, the difference in the data at high temperatures also suggests a considerable phonon softening in PdCrO$_2$ relative to PdCoO$_2$. To address this, we conducted density functional theory calculations of the phonon spectra of PdCoO$_2$, a hypothetical non-magnetic PdCrO$_2$ and PdCrO$_2$ in which we included spin polarization. For the latter, we assumed ferromagnetic order. The calculated structural properties, however, should be dominated by the effect of the local Cr spin polarization and essentially independent from the specific magnetic order. 

The results for heat capacity are shown in Fig. \ref{fig5}b, and the calculated lattice properties for the two materials in Table \ref{ST} in the Appendix.  Due to the different band fillings, the chemical bonding is slightly stronger in PdCoO$_2$ (fully filled Co 3$d$-$t$$_{2g}$ states) than in the fictitious non-magnetic PdCrO$_2$ as reflected in the calculated bulk moduli (see Table \ref{ST}). However, incorporating the Cr spin 3/2 moment in a spin polarized calculation leads to a substantial additional softening (approximately 15$\%$), together with a large expansion of the unit cell volume and strongly improved agreement with the experimental crystal structure. The good agreement with the available experimental data (see table \ref{ST}), in particular for the relative changes between PdCoO$_2$ and magnetic PdCrO$_2$, provides confidence in the calculated properties like phonon spectra (see figures \ref{s1}, \ref{s2}) and the vibrational part of the specific heat and entropy (see Figs. \ref{fig5}b, \ref{fig5}c). We also note that both our measurements and calculations for PdCoO$_2$ agree well with those recently reported in Ref. \cite{Andy2}

Our results show a significant difference between the vibrational high-temperature heat capacities of the two materials, with that of PdCrO$_2$ being larger, in agreement with experiment. The intuitive picture for this is that the existence of the local Cr magnetic moment creates occupancy restrictions on the Cr 3$d$ orbitals, causing them to expand and increasing the cell volume. However, this expansion is relatively soft, lowering the bulk modulus and the Debye temperature. A detailed discussion of the calculations can be found in the Appendix.  

The DFT calculations for the phonon spectrum of spin-polarized PdCrO$_2$ take account of the effect of the Cr moments, but include neither a magnetic phase transition nor a calculation of magnon entropy. They thus allow us to estimate the phonon contribution to the entropy difference between magnetic PdCrO$_2$ and non-magnetic PdCoO$_2$. This phonon entropy difference is plotted as a function of temperature in Fig. \ref{fig5}c, along with the measured entropy difference calculated from the actual heat capacity data in Fig. \ref{fig5}a. Subtracting the phonon contribution isolates the magnetic entropy of PdCrO$_2$, which is seen to be very close to Rln4 at our maximum measurement temperature of 390 K. The very small difference that is seen, with the magnetic entropy not perfectly saturating to Rln4, may be due to small errors in the DFT-calculated phonon entropy value.  Overall, however, the match between experiment and the simple expectation for the entropy from fluctuating spin 3/2 moments is excellent. These fluctuating spins can be expected to scatter the conduction electrons.

Direct empirical comparison of the resistivities of the two compounds confirms coupling between the conduction electrons and the Mott insulating layer in PdCrO$_2$ strongly enhances their scattering rate. In Fig. \ref{fig6} we show the resistivity of the PdCrO$_2$ and PdCoO$_2$ microstructures studied in this project, overlaying the previously-reported results on bulk single crystals from Ref. \cite{24}. The difference between the two, also plotted in Fig. \ref{fig6}, is substantial – larger than the resistivity of PdCoO$_2$. Taken together with the behaviour of the resistivity of PdCrO$_2$ across its magnetic transition, the natural explanation is strong scattering of the conduction electrons from the Pd layers due to their close proximity to the Mott insulating CrO$_2$ layers \cite{27}. However, the situation is not as simple as scattering from fluctuating magnetic moments, as we now discuss. 

\section{Discussion}

\begin{figure}
    \centering
    \includegraphics[width=0.48\textwidth]{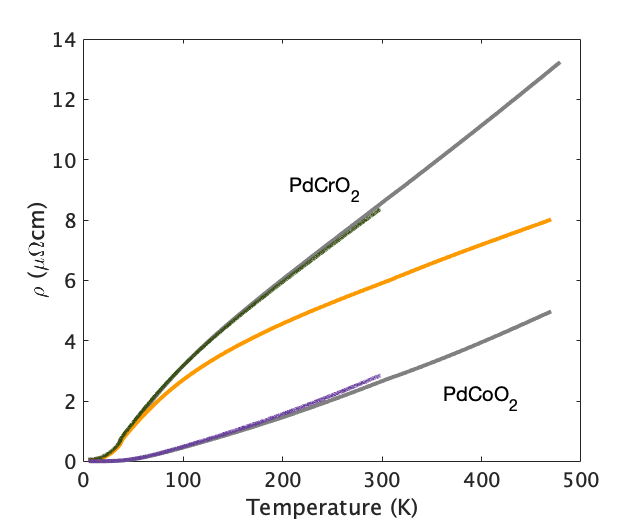}
    \caption{Resistivity of magnetic PdCrO$_2$ and non-magnetic PdCoO$_2$ from the microstructures studied in this work (green and purple lines) and the bulk crystals studied in ref. \cite{24} (grey lines), and the difference between the two (orange).}
    \label{fig6}
\end{figure}

The results presented in Figs. \ref{fig3} and \ref{fig6} suggest a Matthiessen’s rule approach to understanding the resistivity of PdCrO$_2$, namely that $\frac{1}{\tau}$ = $\frac{1}{\tau_{imp}}$ + $\frac{1}{\tau_{ph}}$ + $\frac{1}{\tau_{mag}}$ where $\frac{1}{\tau_{imp}}$, $\frac{1}{\tau_{ph}}$ and $\frac{1}{\tau_{mag}}$ are the scattering rates of electrons from impurities, phonons and magnetic excitations respectively. However, a further issue then needs to be considered.  The Lorenz ratio data shown in Fig. \ref{fig4} b) indicate that the scattering is quasi-elastic above approximately 100 K.  In the electron-phonon problem this is known to lead to $T$-linear resistivity because although the scattering at high temperatures is nearly elastic, the cross-section depends linearly on temperature. Crudely, raising the temperature increases the vibration amplitude of the ions, increasing the scattering cross-section \cite{14}. This can also be viewed as the number of excited phonons in each mode being proportional to temperature. For a spin system, however, this is not the case.  Once the spins are maximally disordered, their scattering cross-section saturates to some temperature-independent value \cite{arxiv}. The magnetic entropy data shown in Fig. \ref{fig5} suggest that this has occurred by 400 K in PdCrO$_2$, yet the magnetic contribution to the resistivity deduced as the orange curve in Fig. \ref{fig6} continues to rise. This suggests that something more subtle is going on in PdCrO$_2$. As well as causing direct scattering from the spin system in the CrO$_2$ layers, the Kondo coupling between the Pd and CrO$_2$ layers must have an additional effect on the resistivity. In the companion paper to this one \cite{theory}, we describe a theoretical solution to the problem, in which the missing high temperature scattering is attributed to fluctuations in the Kondo coupling itself. Although the theory of Ref. \cite{theory} explains the contributions that must be summed to obtain the observed $T$-linear resistivity, it does not provide a first-principles reason for the Planckian value of the scattering rate. Rather, that value must be assumed to be the result of fine tuning. The physical significance of that last finding is not yet fully clear. It at least leaves open the possibility of a bound on the scattering rate that remains to be fully understood. Such a situation would have a deeper signficance than a bound on phonon scattering alone \cite{Andy3}, because it would be a bound on the total scattering rate, even when it had both magnetic and phononic origins.  These considerations motivate further work on the Planckian scattering problem.

\section{Conclusion}

We have performed direct empirical comparisons of the physical properties of the non-magnetic layered metal PdCoO$_2$ and its magnetic counterpart PdCrO$_2$. The use of microstructuring facilitated well-controlled, high-accuracy measurement of the electrical resistivity of the two materials, and of the effect on PdCrO$_2$ of controlled levels of point disorder. Even as the residual resistivity is varied by over a factor of twenty, the $T$-linear resistivity seen at high temperatures retains the same gradient, with the deduced scattering rate within 10$\%$ of the Planckian value. Complementary study of the thermal conductivity and high temperature entropy shows that the resistivity of PdCrO$_2$ is influenced in two ways by the Kondo coupling between its conducting Pd layers and the moments in the CrO$_2$ layers. This slightly surprising observation motivates careful theoretical analysis.  It is particularly striking to realize that the Planckian resistivity is made up of a sum of contributions, none of which is individually $T$-linear. This hints at an overall bound on total scattering rate playing a role in the Planckian problem.

\begin{acknowledgments}

EZ, PHM, SJK, SK, HR and APM thank the Max Planck Society for its support. This work builds on earlier work in collaboration with Veronika Sunko, Phil King, Roderich Moessner, Takashi Oka and Sota Kitamura, all of whom, as well as Steve Kivelson, we thank for useful discussions. Ulrike Nitzsche is acknowledged for technical support.

We thank Olivier Cavani for operating the SIRIUS accelerator. Irradiation experiments performed on the SIRIUS platform
were supported by the French National Network of
Accelerators for Irradiation and Analysis of Molecules
and Materials (EMIR$\&$A) under Project No. EMIR 20-3000. Research in Dresden benefits from the environment provided by the DFG Cluster of Excellence ct.qmat (EXC 2147, project ID 390858940).

 JFMV and DC are supported by faculty startup grants at Cornell University. ET and EB were supported by the European Research Council (ERC) under grant HQMAT (Grant Agreement No. 817799), the Israel-US Binational Science Foundation (BSF), and the Minerva Foundation.

\end{acknowledgments}

\appendix

\section{Supporting information for density functional calculations}

For non-magnetic PdCoO$_2$, the calculated lattice parameters and cell volume are in good agreement with the experimental observation. As often observed empirically in transition metal oxide compounds, the LDA slightly underestimates the bond lengths (and therefore the lattice parameters), whereas GGA leads to slight overestimates. However, an average of both approximations leads to error compensation and a result that is within 1$\%$ of the experimental value (DFT: 122.0 \AA$^3$ vs. Exp.: 123.1 \AA$^3$). This applies also for the calculated bulk modulus (DFT: 221.2 GPa vs. Exp.: 224.0 GPa). This good agreement between the calculated and the experimental lattice properties also provides confidence in the calculated phonon spectra and specific heat.

\begin{figure}
    \includegraphics[width=0.48\textwidth]{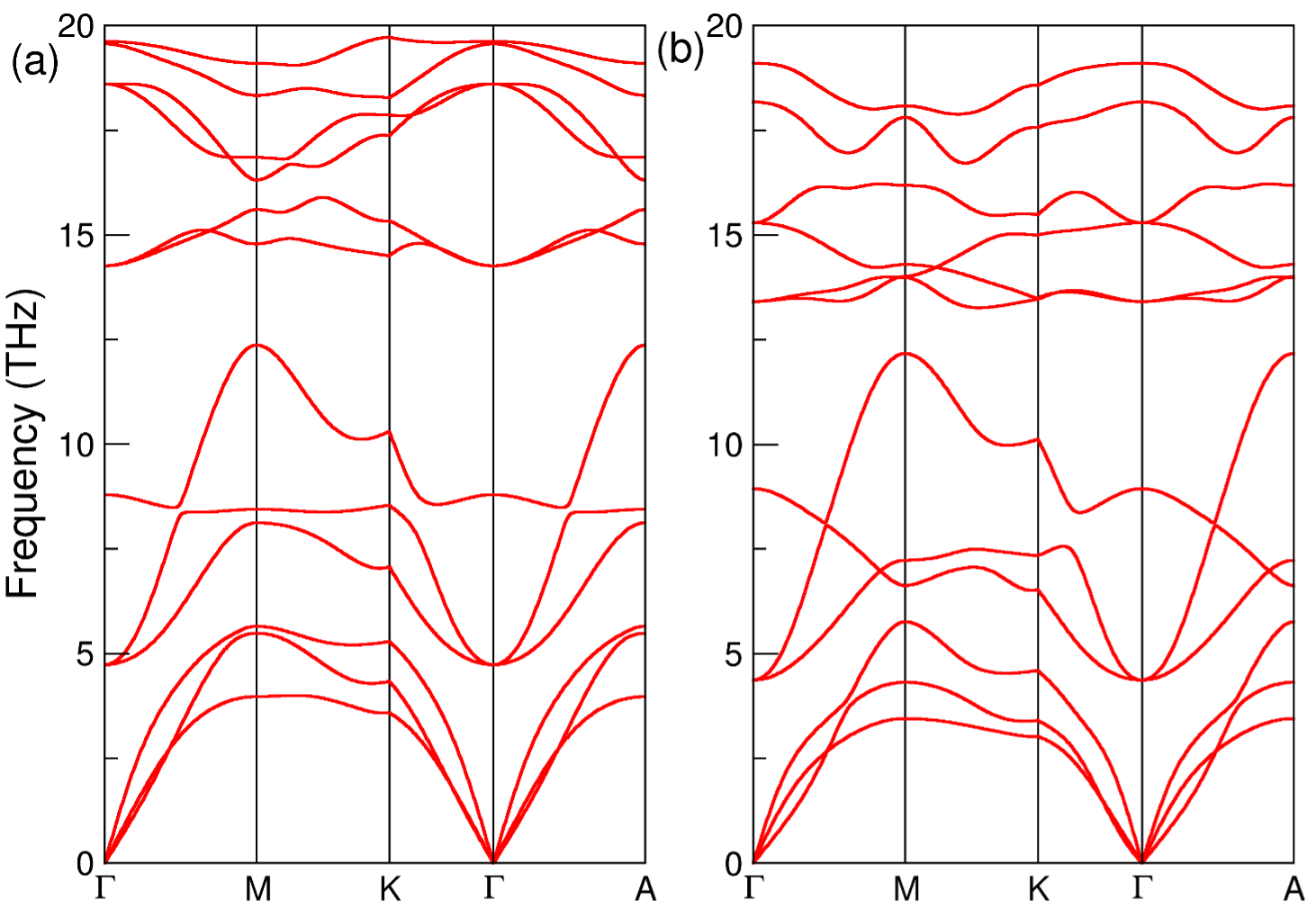}
    \caption{Calculated phonon dispersions along the standard hexagonal band path of (a) PdCoO$_2$ (non-magnetic calculation) and (b) PdCrO$_2$ (spin polarized calculation) applying GGA as the exchange correlation potential. The spectra of both compounds are rather similar, as expected given their structural similarity.  However, the frequencies in the Cr compound are slightly smaller, in particular for the high energy part of the spectrum.}
    \label{s1}
\end{figure}

\begin{figure}
    \includegraphics[width=0.48\textwidth]{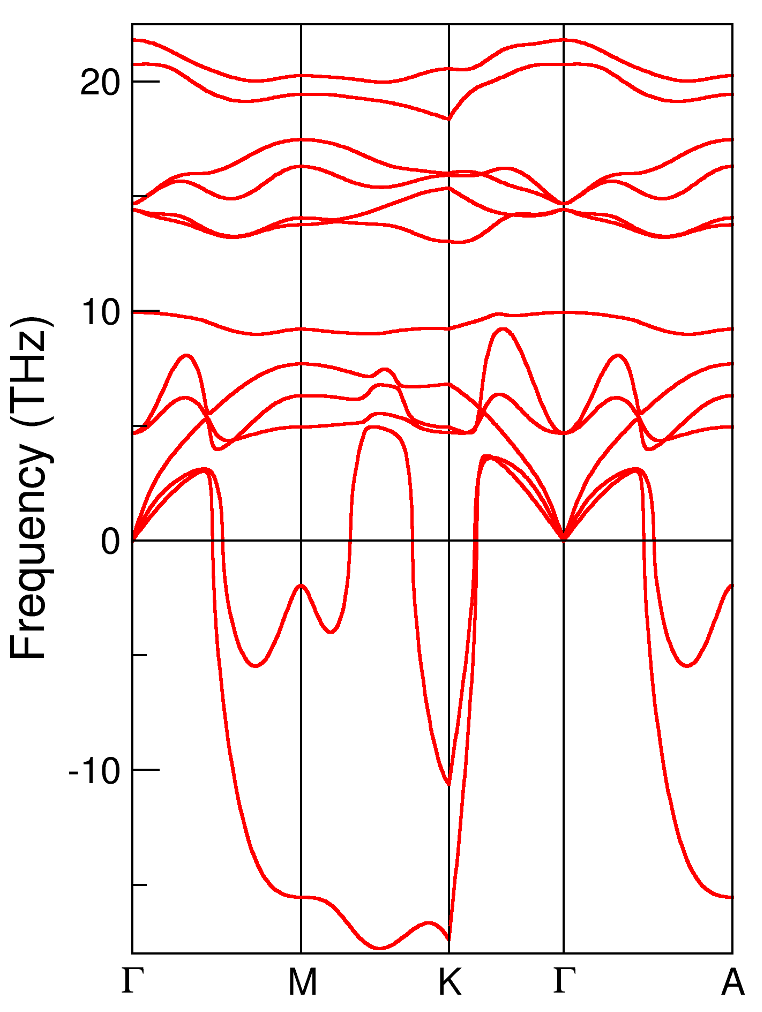}
    \caption{Calculated phonon dispersions of non-spin-polarized PdCrO$_2$ (LDA, a converged GGA calculation could not be obtained). The dispersion of the non-spin-polarized PdCrO$_2$ shows imaginary frequencies, which implies the instability of the lattice without Cr spin polarization.}
    \label{s2}
\end{figure}

\begin{figure}
    \includegraphics[width=0.48\textwidth]{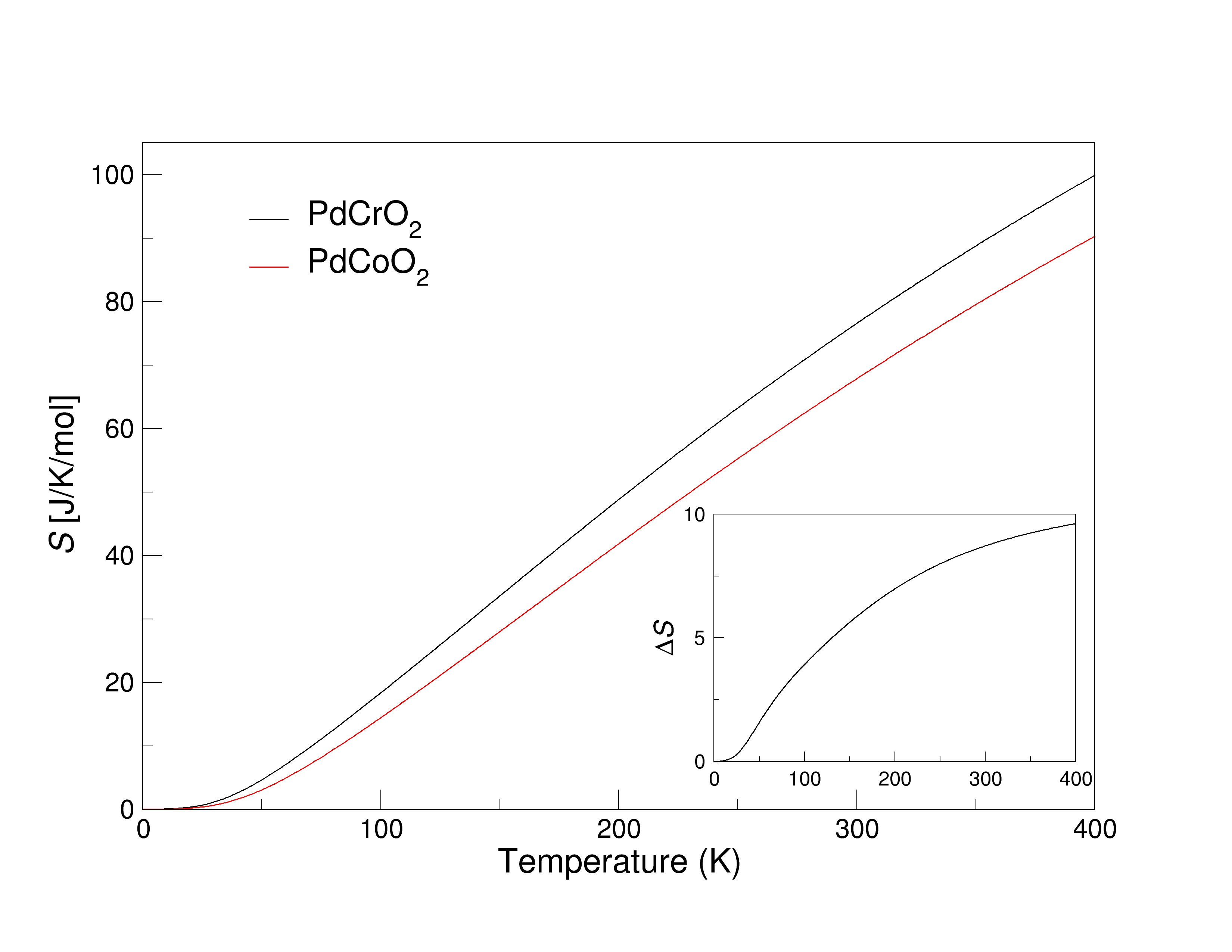}
    \caption{Calculated vibrational entropy of PdCrO$_2$ (black solid line) and PdCoO$_2$ (red solid line) applying GGA for the exchange correlation functional. For PdCrO$_2$ the inclusion of spin polarization was required to obtain a stable lattice. Inset: Calculated vibrational entropy difference $\Delta$$S$ = $S$$_{\mathrm {PdCrO_2}}$ - $S$$_{\mathrm {PdCoO_2}}$ between PdCrO$_2$ and PdCoO$_2$. }
    \label{s3}
\end{figure}

\begin{table}
\caption{\label{tab:table2} Calculated lattice parameters and bulk moduli of PdTO$_2$ (T = Co, Cr) applying different density functionals (LDA vs. GGA). The bulk moduli were obtained by fitting the total energies of different volumes to the Birch-Murnaghan equation. Our results are compared to theoretical and experimental literature data. “NM” refers to non-magnetic calculations, “SP” to calculations including spin polarization, in the present case with ferromagnetic order. The dominant change compared to the non-magnetic case, however, originates from the local spin polarization and not from the specific type of inter-site order.}

\begin{ruledtabular}
\begin{tabular}{cccccc}
 Compound & method & V$_0$ [\AA$^3$] & a [\AA] & c/a & B [GPa]\\
\hline
PdCoO$_2$ (NM) & GGA & 126.53 & 2.86 & 6.23 & 194.51\\
 & LDA	&117.43	&2.79	&6.27	&247.82\\

ref. \cite{SS1}	&GGA& 	128.32	&2.86	&6.31&	206.86\\
	&LDA& 	119.02	&2.79	&6.35&	264.75\\

 ref. \cite{SS2}&	Exp.& 	123.06	&2.83&	6.27&	224.00\\

 PdCrO$_2$ (NM)	&GGA	&126.60&	2.76&	6.94&	180.82\\
 	&LDA&	117.19&	2.68	&7.02	&238.85\\

  PdCrO$_2$ (SP)	&GGA	&138.65&	2.97&	6.09&	167.40\\
  &LDA	&127.89&	2.88&	6.19&	194.50\\

  ref. \cite{27}	&Exp. 	&133.82	&2.92	&6.19	&-
  \label{ST}
\end{tabular}
\end{ruledtabular}
\end{table}

In contrast to PdCoO$_2$, a (hypothetical) non-magnetic calculation for PdCrO$_2$ leads to a strong underestimate of the calculated lattice parameters and cell volume. Whereas the calculated cell volume of both sister compounds is essentially identical, the experimental value for the Cr compound is about 10$\%$ larger. This clearly indicates the importance of the local spin polarization for Cr$^{3+}$ for a more accurate description of the lattice properties. Indeed, a spin polarized calculation (with ferromagnetic inter-site order) yields an increase of the (LDA-GGA averaged) cell volume of about 10$\%$, resulting in very good agreement with the experimental value (DFT: 133.2 \AA$^3$ vs. Exp.: 133.8 \AA$^3$). This is directly connected with a strong softening for the bulk modulus by about 15$\%$ compared to the non-magnetic calculation.  There is no known experimental value for the bulk modulus of PdCrO$_2$; we predict B = 181 GPa from our spin-polarized calculations. This is more than 20$\%$ smaller than that of PdCoO$_2$. Since the energy scale for the on-site spin polarization is about an order of magnitude larger than that of the inter-site coupling, we expect the specific inter-site magnetic order to have only a small influence on lattice parameters and bulk modulus. 
From the very good agreement between the calculated lattice properties (lattice parameters, cell volume, bulk modulus) for both compounds with the available experimental data we expect also a good agreement for other related lattice properties like the phonon spectra and the derived vibrational parts of the specific heat and the entropy (compare figure \ref{fig5}b main text and figures \ref{s1} - \ref{s3} of the Appendix). 

\bibliography{apssamp}

\end{document}